\documentstyle[prl,aps]{revtex}

\def\LQCD{\Lambda_{\rm QCD}}
\def\q{{\bf q}}
\def\d{\partial}
\def\permea{\lambda}

\begin{document}

\twocolumn[\hsize\textwidth\columnwidth\hsize\csname@twocolumnfalse\endcsname
\preprint{CU-TP-xxx}
\title{Asymptotic Deconfinement in High-Density QCD}
\author{D.H.~Rischke,$^{1,4}$ D.T.~Son,$^{2,4}$ and M.A.~Stephanov$^{3,4}$}
\address{$^1$Nuclear Theory Group, Brookhaven National
Laboratory, Upton, New York 11973}
\address{$^2$Physics Department, Columbia University, New York, New York 10027}
\address{$^3$Department of Physics, University of Illinois, Chicago,
Illinois 60607-7059}
\address{$^4$RIKEN-BNL Research Center, Brookhaven National Laboratory,
Upton, New York 11973}
%\date{November 2000}
\maketitle

\begin{abstract}

We discuss QCD with two light flavors at large baryon chemical
potential $\mu$.  Color superconductivity leads to partial breaking of
the color SU(3) group.  We show that the infrared physics is governed
by the gluodynamics of the remaining SU(2) group with an exponentially
soft confinement scale $\LQCD'\sim\Delta\exp(-a\mu/(g\Delta))$, where
$\Delta\ll\mu$ is the superconducting gap, $g$ is the strong coupling,
and $a=2\sqrt{2}\pi/11$.  We estimate that, at moderate baryon
densities, $\LQCD'$ is ${\cal O}$(10 MeV) or smaller.  The confinement
radius increases exponentially with density, leading to ``asymptotic
deconfinement.''  The velocity of the SU(2) gluons is small due to the
large dielectric constant of the medium.

\end{abstract}
\vskip 2pc]

%\section{Introduction}

{\em Introduction}.---%
Soon after the discovery of asymptotic freedom in QCD \cite{AF} a hypothesis
was put forward that, at high baryon densities, quarks (which are
normally confined in hadrons by strong forces) are liberated, i.e.,
nuclear matter transforms into deconfined quark matter
\cite{CollinsPerry}.  In recent years, our knowledge of dense quark
matter has considerably expanded.  We now understand that, in reality,
dense matter shows more intricate features than in the original
picture of \cite{CollinsPerry}.  In particular, quark matter at high
densities exhibits the phenomenon of color superconductivity
\cite{CS,CS1}, which determines the symmetry of the ground state
and the infrared dynamics.

The number of
%relevant
light quark flavors $N_f$ turns out to play a
crucial role.
%In most applications, e.g. at densities characteristic
%of neutron stars,
The simplest case is $N_f=2$, where up and down quarks are massless
and other quarks are neglected.
%up and down quarks are relevant.
The following picture emerges in perturbation theory, as well as
in instanton-inspired models.  The condensation of color
antitriplet up-down diquarks breaks the color SU(3) down to an
SU(2) subgroup.  Thus, five of the original eight gluons
acquire ``masses'' by the Meissner effect \cite{CS1,Rischke2fl},
similar to the Higgs mechanism.  The remaining three gluons are
massless (perturbatively).  Because of Cooper pairing, the
spectrum of quark excitations carrying SU(2)
color charge has a gap $\Delta$.  
%At small coupling (large
%densities), $\Delta/\mu$ is exponentially small.
%this gap is exponentially smaller than $\mu$.  

In order to understand the physics below the energy scale $\Delta$ we
must examine the pure gluodynamics in the remaining unbroken SU(2)
sector.  As we shall see, the process of high-density
``deconfinement'' is quite nontrivial in this case: the quarks are
{\em always} confined (assuming that SU(2) Yang-Mills theory confines),
however, the confinement radius grows {\em
exponentially} with increasing density. We shall also see that,
at 
%D the 
scales much shorter than the confinement radius, the 
%D gluodynamics
dynamics
of the SU(2) gluons
is similar to electrodynamics in a dielectric medium with
large refraction index.

{\em The effective Lagrangian.}---%
%
%
%
%M% 
Below the scale $\Delta$, we expect that the heavy (gapped) degrees of
freedom decouple and the remaining fields can be described by
a local effective Lagrangian.  
The absence of quarks carrying SU(2) charges below
$\Delta$ implies that the medium is transparent to the
SU(2) gluons: there is no Debye screening and Meissner
effect for these gluons.  Mathematically, the polarization tensor
$\Pi^{\mu\nu}_{ab}(q)$ vanishes at $q=0$, which can be checked by a
direct calculation of $\Pi$ at small $q$, as was done in Ref.\
\cite{Rischke2fl}.  The absence of Debye screening means that a static
color charge inserted into the medium cannot be completely screened as
it is in hot plasmas.  This is easy to understand since all quarks carrying
SU(2) color are bound into SU(2) singlet Cooper pairs.
Analogously, the Meissner effect is absent because
the superconducting currents, which
are coherent motions of the condensate, cannot screen the magnetic
field, since the condensate is SU(2) neutral.
Thus, at first sight, it might seem that the quarks in the medium have 
no effect on the gluon
effective Lagrangian, which must be simply $L=-F_{\mu\nu}^2/(4g^2)$, i.e., the
SU(2) Yang-Mills Lagrangian with the coupling $g$ matching the running
coupling in the original theory at the scale $\Delta$ \cite{wrong}.
However, a
closer look shows that the situation is somewhat more complicated and,
in fact, more interesting.

Although a static SU(2) charge cannot be completely
%M
Debye screened by SU(2) neutral Cooper pairs,
it can still be {\em partially} screened if the
medium is polarizable, i.e., if it
has a {\em dielectric constant} $\epsilon$ different from
unity.  If $\epsilon>1$, then the Coulomb potential between two static
color charges is $g^2/(\epsilon r)$; i.e., the gauge coupling is
effectively reduced by a factor of $\epsilon^{1/2}$.
 As explained in more detail below, this is exactly the
situation 
%occurring 
in the color-superconducting phase.
Analogously the medium can, in principle, have a magnetic
permeability $\permea\neq1$. (We denote the permeability by
$\permea$ instead of the more
%D standard 
common $\mu$, since the latter
symbol is already used 
for the chemical potential.)  The dynamics
of gluons is thus modified by the dielectric constant and the
magnetic permeability of the medium.  Hence, one needs to develop
%D a 
the theory of ``gluodynamics of continuous media'', which, as far as
we know, has never been encountered before.  This theory, in
contrast to its U(1) counterpart (the electrodynamics of
continuous media), is an {\em interacting} theory.

Fortunately, even without explicit calculation
we can already write the effective Lagrangian of SU(2) gluons
from rather general arguments.  It should satisfy the requirements of
locality (since the quarks that have been integrated out have gaps)
and gauge invariance.  It does not need to be Lorentz invariant, since
this invariance is already violated by the presence of the high-density
medium, but it should be rotationally invariant and conserve parity.
Thus the effective action at the scale $\Delta$
must have the following form
\begin{equation}
  S_{\rm eff} = {1\over g^2}\int\!d^4x\,\biggl({\epsilon \over 2}
  {\bf E}^a\cdot {\bf E}^a - {1\over2\permea}{\bf B}^a\cdot {\bf B}^a
  \biggr) \, ,
  \label{Leff}
\end{equation}
where $E^a_i\equiv F_{0i}^a$ and
$B^a_i\equiv{1\over2}\epsilon_{ijk}F^a_{jk}$.  Higher-order
corrections (in powers of fields and derivatives) to (\ref{Leff})
are irrelevant for the infrared physics and have been neglected.
The constants $\epsilon$ and $\permea$ in Eq.\ (\ref{Leff}) have
the meaning of the the dielectric constant and the magnetic
permeability in the regime where the gluon fields are linear.  
%In
%particular, $v=1/\sqrt{\epsilon\permea}$ is the speed of gluons in
%this regime. 
%Implicitly, the ultraviolet cutoff in (\ref{Leff})
%is $\Delta$.
The speed of gluons in this regime is $v=1/\sqrt{\epsilon\permea}$.

As we shall see, $\epsilon\permea\gg1$. This should be
contrasted with the standard picture of 
%D
the vacuum as a dielectric,
in which case $\epsilon_0\permea_0=1$ by Lorentz invariance.
In other words, vacuum polarization effects (such as 
gluon, ghost, 
%D loops 
and $\mu=0$ quark loops) 
%D 
in Eq.\ (\ref{Leff})
%D can be 
have been absorbed into
the running of $g^2$.
%D , which is implicitly done in Eq.(\ref{Leff}).
The constants $\epsilon$ and $\permea$ are defined relative 
to those of 
%D
the vacuum.

% To be precise, the the constants $\epsilon$ and $\lambda$ are
% defined relative to those of vacuum, $\epsilon_0$ and $\mu_0$.
% Since vacuum is polarizable, one should specify the energy scale
% at which $\epsilon_0$ and $\mu_0$ are computed. The most
% convenient choice is to set $\epsilon_0=\mu_0=1$ at the Debye
% screening scale $g\mu$. With this choice, one can forget all
% vacuum effects (in particular, gluon and ghost loops) and compute
% only effects due to the medium.

One can give a parametrically correct estimate of $\epsilon$ using a
crude model, in which the Cooper pairs are represented by classical
oscillators with the spring constant $k$. The characteristic scale for
a Cooper pair ``oscillator'' is $\Delta$; thus one can estimate
$k\sim\Delta^3$.  The polarizability of each oscillator is $g^2/k$.
Multiplying by the density of the Cooper pairs, $\mu^2\Delta$, one
finds $\epsilon-1\sim g^2\mu^2/\Delta^2$. This estimate is
corroborated by the explicit calculation, presented 
%D in the latter part of the 
later in this paper, where we obtain
\begin{eqnarray}
  \epsilon &=& 1 + \kappa = 1 + {g^2\mu^2\over18\pi^2\Delta^2}
    \, , \label{epsilon} \\
  \permea &=& 1 \, . \label{mu}
\end{eqnarray}

At high densities, the gap $\Delta$ is exponentially
suppressed compared to the chemical potential $\mu$ \cite{Son},
\begin{equation}
  \Delta = b \mu g^{-5}e^{-c/g}, \qquad c = {3\pi^2\over\sqrt{2}} \, ,
  \label{Delta}
\end{equation}
where $g$ is the gauge coupling at the scale $\mu$, and $b$ is some
numerical constant.  According to Eq.\ (\ref{epsilon}), $\kappa\gg1$ and
we can write
\begin{equation}
  \epsilon \approx {g^2\mu^2\over18\pi^2\Delta^2} \gg 1 \, ,
  \label{epsilon2}
\end{equation}
which means that the dielectric constant of the medium is very large.
Hence, the Coulomb potential between SU(2) color charges is greatly
reduced.  This can be interpreted as a consequence of the fact that the
Cooper pairs have large size (of order $1/\Delta$) and so are easy to
polarize.  The magnetic permeability, in contrast, remains close
to 1 due to the absence of mechanisms that would strongly screen the
magnetic field.

{\em The scale of confinement}.---%
Once the effective Lagrangian (\ref{Leff}) is obtained, one
can use it to investigate the infrared dynamics of the gluons.
One notices that (\ref{Leff}) possesses a modified Lorentz symmetry in
which the speed of light $c=1$ is replaced by
\begin{equation}
  v = {1\over\sqrt{\epsilon}} \, .
  \label{v}
\end{equation}
One can make this symmetry manifest by rescaling the time,
the field $A_0$ and the coupling in Eq.\ (\ref{Leff}),
\begin{eqnarray}
%   x^0 & \to & x^{0\prime} = {x^0\over\sqrt{\epsilon}} \, , \nonumber \\
%   A_0^a & \to & A_0^{a\prime} = \epsilon^{3/4} A^a_0 \, , \nonumber \\
%   A_i^a & \to & A_i^{a\prime} = \epsilon^{1/4} A^a_i \, , \nonumber \\
%   g & \to & g' = {g\over\epsilon^{1/4}} \, . \label{rescale}
x^{0\prime} = {x^0\over\sqrt{\epsilon}} \, , \quad
A_0^{a\prime} = \sqrt\epsilon A^a_0 \, , \quad
g' = {g\over\epsilon^{1/4}} \, . \label{rescale}
\end{eqnarray}
After the rescaling (\ref{rescale}), the action (\ref{Leff})
assumes the familiar Lorentz-invariant form in the new coordinates,
\begin{equation}
  S = -{1\over4g^{\prime2}}
\int\!d^4x'\, F_{\mu\nu}^{a\prime} F_{\mu\nu}^{a\prime} \, ,
  \label{S'}
\end{equation}
where
\begin{equation}
  F_{\mu\nu}^{a\prime} = \d_\mu' A_\nu^{a\prime} - \d_\nu A_\mu^{a\prime}
     + f^{abc} A_\mu^{b\prime} A_\nu^{c\prime}\ .
  \label{F'}
\end{equation}
The coupling in the action (\ref{S'}) is not $g$ but $g'$
which is smaller by a factor of $\epsilon^{1/4}$.  This means that the
small parameter that controls the perturbative expansion in the theory
(\ref{Leff}) is not $\alpha_s=g^2/(4\pi)$ but rather
\begin{equation}
  \alpha_s' = {g^2\over4\pi\sqrt{\epsilon}} \, ,
  \label{alphas'}
\end{equation}
which is much smaller than $\alpha_s$, since $\epsilon$ is large.

Another way to derive Eq.\ (\ref{alphas'}) is by restoring 
%the factors of 
$\hbar$ and $c$ in the expression for the strong coupling constant
$\alpha_s$, which is given by $g^2/(4\pi\hbar c)$ in the vacuum.  In our
dielectric medium, the Coulomb potential between two static charges
separated by $r$ is $g^2/(\epsilon r)$.  Thus, we have to
replace $g^2$ by $g^2_{\rm eff}=g^2/\epsilon$.  The velocity of light
$c$ also needs to be replaced by the velocity of gluons $v$.  This
gives
\begin{equation}
  {g_{\rm eff}^2\over 4\pi \hbar v} = {g^2\over4\pi\sqrt{\epsilon}} \, ,
  \label{alphaeff}
\end{equation}
since $\hbar=1$ in our unit system.  Equation (\ref{alphaeff}) coincides
with $\alpha_s'$ in Eq.\ (\ref{alphas'}), as one expects.

Using Eq.\ (\ref{epsilon2}), one can express the coupling $\alpha_s'$
in terms of the gap $\Delta$,
\begin{equation}
  \alpha_s' = {3\over 2\sqrt{2}} {g\Delta\over\mu} \, .
  \label{alphaDelta}
\end{equation}
Equations (\ref{alphas'}) and (\ref{alphaDelta}) define the coupling
in our effective theory at the matching scale with the original
microscopic theory, i.e., at the scale $\Delta$.  The coupling
increases logarithmically
as one moves to lower energies, since  
%D the SU(2) pure
pure SU(2)
Yang-Mills theory is asymptotically free.  This coupling becomes large
at the confinement scale $\LQCD'$, which is the mass scale of SU(2)
glueballs. The spectrum of these glueballs is known
from lattice studies of SU(2) Yang-Mills theory
\cite{MichaelTeper87-88}, except that the role of the speed of light
is now played by $v$ [Eq.\ (\ref{v})].
Since $\alpha_s'$ is tiny [because 
%D of 
$\Delta/\mu\ll1$ in 
Eq.\ (\ref{alphaDelta})] it takes long to grow, and
the scale $\LQCD'$ is thus very small.
%, since the coupling
%$\alpha_s'$ at the $\Delta$ scale is tiny, Eq.\ (\ref{alphaDelta}).
Using the one-loop beta function, one can estimate
\begin{equation}
  \LQCD' \sim \Delta \exp\biggl(-{2\pi\over\beta_0\alpha_s'}\biggr)
  \sim \Delta \exp\biggl(-{2\sqrt{2}\pi\over11}{\mu\over g\Delta}
  \biggr) \, ,
  \label{LQCD'}
\end{equation}
where $\beta_0$ is the first coefficient in the beta function and is
equal to $22/3$ in SU(2) gluodynamics.

We can draw a few immediate conclusions from Eq.\ (\ref{LQCD'}).
First, $\LQCD'$ depends very sensitively on the gap $\Delta$, in
particular on the numerical value of the constant $b$ in Eq.\
(\ref{Delta}).  Unfortunately, the latter is not exactly known.  The
uncertainty in the value of the gap $\Delta$ translates into a huge
variation of $\LQCD'$.  For example, if one uses
%the value
\begin{equation}
  b=512\pi^4 \, ,
  \label{blarge}
\end{equation}
which is obtained by solving the one-loop gap equation where the
exchanged gluon propagator is replaced by the hard dense loop (HDL) expression
\cite{SchaeferWilczekPisarskiRischke}, then with $\LQCD=200$ MeV we
find $\LQCD'\sim10$ MeV at $\mu=600$ MeV.  However, if we use
\begin{equation}
  b=512\pi^4\exp\biggl(-{\pi^2+4\over8}\biggr) \, ,
  \label{bsmall}
\end{equation}
which is obtained if one assumes the Bardeen-Cooper-Schrieffer ratio 
between the critical
temperature $T_c$ and the gap $\Delta$, and computes $T_c$ by taking into
account the fermion wave-function renormalization \cite{Rockefeller},
then, at the same chemical potential, $\LQCD'$ is reduced to a mere 0.3
keV.  Regretfully, neither Eq.\ (\ref{blarge}) nor (\ref{bsmall})
seems to be entirely correct, since there are physical effects that
they do not take into account (e.g., the Meissner effect).
%, although
%Eq.\ (\ref{bsmall}) takes into account more physics than
%(\ref{blarge}).
Clearly, any attempt to give even the roughest
numerical estimate for $\LQCD'$ requires an accurate determination of
the gap $\Delta$.  It has been argued that to compute $\Delta$ one
needs a better understanding of the issue of gauge invariance,
the finite fermion lifetime, and the running of the coupling
\cite{RajagopalShuster_etal}.
Regardless of all these uncertainties, the
exponential dependence of $\LQCD'$ on $\mu/\Delta$ makes it safe to
predict that, even at moderate values of $\mu$, the confinement scale
$\LQCD'$ is very small, much smaller than $\LQCD$.

Second, as the density, i.e., $\mu$, 
is increased, $\LQCD'$ vanishes exponentially
fast due to the factor $\mu/\Delta$ in the exponent.
%, and vanishes at
%$\mu\to\infty$.
We arrive at the following picture of how
deconfinement occurs at large densities.
Strictly speaking, at any given 
density, 
%value of the chemical potential,
the theory is confined.  However, the
confinement radius $1/\LQCD'$
grows exponentially as the density is increased.
Therefore, if one looks at the physics at some large, but fixed,
distance scale, there is a crossover density when
effectively the color degrees of freedom become deconfined at that
scale.  We call this phenomenon ``asymptotic deconfinement.''

{\em The computation of $\epsilon$ and $\permea$}.---%
To find $\epsilon$ and $\permea$, one has to calculate
 (\ref{Leff}) by integrating out the quark degrees of freedom in the
QCD Lagrangian. This amounts to
computing the one-loop polarization operator $\Pi(q)$ and the gluon
vertices $\Gamma_3(q_1, q_2)$, $\Gamma_4(q_1,q_2,q_3)$, etc.  This
procedure is the same as the one giving rise to the hard thermal loop
(HTL) and hard dense loop effective actions \cite{HTL}.  The
situation here is simpler than in the HTL and HDL cases:
in the regime where all gluon
momenta $q$ are much smaller than $\Delta$, the functions
$\Pi$ and $\Gamma$ can
be expanded in powers of $q$, yielding a local effective Lagrangian.
(In contrast, the HTL and HDL actions are non-local, since the
fermions do not have gaps.)  The gauge invariance of the effective
Lagrangian greatly simplifies our task: in order to know $\epsilon$
and $\permea$, one needs to compute only the polarization tensor $\Pi$
of the SU(2) gluons.  The leading contribution at large density comes
from the superconducting quark loop.
 The detailed calculation of 
$\Pi$
%the
%polarization tensor 
was done in Ref.\ \cite{Rischke2fl}. From Eq.\
(99a) of that paper one can derive the following expression for
$\Pi^{00}_{ab}(q_0,{\bf q})$, $a,b=1,2,3$:
\begin{eqnarray}
  \Pi^{00}_{ab}(q_0, \q) &=& - \delta_{ab}{g^2\mu^2\over\pi^2}
  {\Delta\over q}
  \int\limits_0^\infty\! dz\!\!\int\limits_0^{q/2\Delta}\! dy \,
  \biggl( 1 - {z^2-y^2+1 \over u_+ u_-} \biggr) \nonumber\\
   & \!\!\!\!\!\!\!\! \times &\!\!\!\!\!\!
  \biggl({1\over u_+ + u_- + q_0/\Delta} +
  {1\over u_+ + u_- - q_0 / \Delta}\biggr)
  \label{Pi00}
  \, ,
\end{eqnarray}
where $u_\pm = \sqrt{(z\pm y)^2+1}$, and we assume that $q_0$ and
$q\equiv|{\bf q}|$ are much smaller than $\mu$ so that the dominant
contribution comes only from particles near the Fermi surface.  
%D
Physically, the first term in parentheses is
the probability to excite a quark-hole pair through an SU(2)
gluon, and the last
term contains the corresponding energy denominators for
such an excitation.
%D
For $|q_0|>2\Delta$, one should replace $q_0$ by $q_0+i\epsilon$.
Expanding Eq.\ (\ref{Pi00}) to quadratic order in $q_0$ and $q$ around
$q_0=q=0$, one finds
\begin{equation}
  \Pi^{00}_{ab}(q_0, \q) = -\kappa\,  q^2\delta_{ab} \, ,
\end{equation}
where
\begin{equation}
  \kappa  = {g^2\mu^2\over18\pi^2\Delta^2} \, .
  \label{kappa}
\end{equation}
The appearance of 
%D
a factor $\mu^2$ is due to
the fact that the loop integral is dominated by the momentum region
near the Fermi surface, whose area is proportional to $\mu^2$.
Similarly, one obtains from Eq.\ (99b) of \cite{Rischke2fl} the following:
\begin{equation}
  \Pi^{0i}_{ab}(q_0, \q) = -\kappa\,  q^0 q^i\delta_{ab} \, .
\end{equation}
The computation of $\Pi^{ij}_{ab}(q_0,\q)$ is facilitated by writing
\begin{eqnarray}
  \Pi(q_0,\q) & \equiv &[\Pi(q_0, \q) - \Pi(0, \q)] +
  [ \Pi(0,\q) - \Pi_{\rm HDL}(0,\q) ] \nonumber \\
  & + &  \Pi_{\rm HDL}(0,\q) \, ,
  \label{mani}
\end{eqnarray}
where $\Pi_{\rm HDL}$ is the standard HDL gluon
self-energy, which vanishes for $q_0=0$.  The term $\Pi(0,\q) -
\Pi_{\rm HDL}(0,\q)$ can be shown to be of order ${\cal
O}(\Delta^2)$ and thus negligible compared to the first term
$\Pi(q_0, \q) - \Pi(0, \q)$.  The reason for the manipulation
(\ref{mani}) is to get rid of the antiparticle contributions in this
term.  With Eq.\ (99c) of \cite{Rischke2fl}, the result for
$\Pi^{ij}_{ab}$ can be written analogously to Eq.\ (\ref{Pi00}) as
\begin{eqnarray}
  \lefteqn{
  \Pi^{ij}_{ab}(q_0, \q) = - \delta_{ab}{g^2\mu^2\over\pi^2}
  {\Delta\over q}
  \int\limits_0^\infty\! dz\!\!\int\limits_0^{q/2\Delta}\! dy
  \int\limits_0^{2\pi} \! {d\varphi\over 2\pi}
  \, {\hat k}^i {\hat k}^j} \nonumber \\
  & \times &
  \biggl( 1 - {z^2-y^2-1 \over u_+ u_-} \biggr) \label{Piij} \\
  & \times &
  \biggl({1\over u_+ + u_- + q_0/\Delta} +
  {1\over u_+ + u_- - q_0 / \Delta}
  - {2 \over u_+ + u_-} \biggr)\nonumber
    \, ,
\end{eqnarray}
where $\hat{\bf k}=(\sin\theta\cos\varphi,\ \sin\theta\sin\varphi,\
\cos\theta)$, and $\cos\theta=2y\Delta/q$.  Expanding to quadratic
order in $q_0$ and $q$, one finds
\begin{equation}
  \Pi^{ij}_{ab}(q_0, \q) = -\kappa\,  q_0^2 \delta^{ij}\delta_{ab} \, .
\end{equation}
Note that (i) the polarization tensor satisfies current conservation,
$q_\mu\Pi^{\mu\nu}_{ab}=0$;  (ii) at this order,
there is no spatial transverse
contribution $\sim q^2\delta^{ij}-q^iq^j$ to $\Pi^{ij}$, although such
a term is not forbidden by the symmetries.
%At the one-loop level, this transverse
%structure emerges only at the order $q^4$ in the expansion over
%$q/\Delta$.

After the quark loop has been integrated out, the quadratic term in
the effective Lagrangian becomes
$A^{\mu}(-q)[D^{-1}_{\mu\nu}(q)+\Pi_{\mu\nu}(q)]A^\nu(q)/(2g^2)$, where
$D$ is the bare gluon propagator.  Comparing with the
quadratic terms in Eq.\ (\ref{Leff}), we obtain Eqs.\ (\ref{epsilon})
and (\ref{mu}).
%D Trivialiyt of
The fact that $\lambda = 1$ [Eq.\ (\ref{mu})] is 
due to the absence of the spatial transverse term
in $\Pi^{ij}$.  

%We have not discussed other low energy excitations in high density
%2-flavor QCD. The unpaired quark of the third color gives rise to a
%fermion (isospin doublet) mode carrying baryon charge. This mode is neutral
%from the point of view of the SU(2) gluons and does not affect the
%picture we described.  There is also a light pseudoscalar isoscalar
%mode, similar to the $\eta$ meson, which can mix with the pseudoscalar
%glueball. This mode is also neutral and it acquires a small mass due
%to the anomalous breaking of the global $U(1)_A$ symmetry. This mass is
%suppressed by a {\em power} of $\mu$, while, as we have seen, the
%glueball masses are much (i.e., exponentially) smaller at large $\mu$.
%It is also interesting to note that
%since the gap in the SU(2) colored quark spectrum, $\Delta$,
%is much larger than $\LQCD'$, the spectrum of mesons made of these
%quarks resembles that of heavy quarkonia.

%{\em Final remarks}.---%
In our analysis above we have neglected other low energy
excitations: the unpaired fermions of the third color and the
pseudoscalar isoscalar mode similar to the $\eta$ meson.  This is
justified because they are colorless with respect to the unbroken
SU(2)$_c$ gluons. It is also interesting to note that
since the gap in the SU(2) colored quark spectrum, $\Delta$,
is much larger than $\LQCD'$, the spectrum of mesons made of these
quarks must resemble  that of heavy quarkonia.

The asymptotic deconfinement phenomenon is
%M%
specific to the case $N_f=2$. At sufficiently high densities
when, effectively, $N_f=3$, the ground state is the color-flavor 
%D locking 
locked (CFL) state:
%M
the color symmetry is broken completely \cite{CFL} and all
gluons are screened, both electrically and magnetically
\cite{Rischke3fl}.
The low energy modes in the CFL phase are
the Goldstone modes described by a chiral effective
Lagrangian \cite{CasalbuoniGatto,inverse}.
Between the two regimes, when one has strange quark matter
below the ``unlocking'' phase transition \cite{unlocking}, the
system might be an anisotropic dielectric if 
%D the 
an $ss$ condensate
breaks rotational invariance. Another interesting regime with asymptotic
deconfinement is 
%D the high isospin density regime 
that of high isospin density \cite{isospin}, where the
gauge group of the 
%pure Yang-Mills theory 
gluodynamics of continuous media
is SU(3).

%So far, we have considered an idealized QCD with two light quarks.
%at finite baryon densities.
%At sufficiently large $\mu$,
%the number of
%relevant
%light quarks becomes $N_f=3$,
%and the ground state is the color-flavor-locking (CFL) state where
%the SU(3) color symmetry is broken completely \cite{CFL}.
%Correspondingly, the Debye and Meissner masses do not vanish in the
%limit of constant fields $q\to0$ \cite{Rischke3fl}.  There is no pure
%Yang-Mills gauge sector, and hence no glueballs, in the infrared.
%Instead, the only light modes are the Goldstone bosons which arise
%from the spontaneous breaking of global symmetries \cite{CFL}.  The
%dynamics of these modes is described by a non-linear sigma model
%\cite{CasalbuoniGatto}, whose parameters can be determined by a
%perturbative computation \cite{inverse}.
%In the real world the formulas derived in this paper
%are valid for $\mu$ below the ``unlocking'' phase transition

%Another regime of QCD with asymptotic deconfinement is that of
%large isospin density \cite{isospin}.  In this regime the gauge
%group SU(3) remains unbroken by the condensate, which is color
%neutral.  The physics below the gap scale $\Delta$ is described by
%pure SU(3) gluodynamics of the type (\ref{Leff}), and asymptotic
%deconfinement occurs at large isospin chemical potential.

The authors are indebted to D.B.~Kaplan, R.D.~Pisarski and
A.R.~Zhitnitsky for stimulating discussions.  We thank RIKEN,
Brookhaven National Laboratory, and U.S.\ Department of Energy
%[DE-AC02-98CH10886]
for providing the facilities essential for the completion of this
work.  The work of D.T.S. is supported, in part, by
% DOE Grant No.\ DE/FG02-92ER40699 and
a DOE OJI grant.

\end{document}